\documentclass[aps,prr,twocolumn,showpacs,floatfix,superscriptaddress,floatfix]{revtex4-2}
\usepackage{natbib}
\usepackage{bm}
\usepackage[normalem]{ulem}
\usepackage{lipsum}
\newcommand{\ex}[1]{\langle#1\rangle}

\PassOptionsToPackage{table,dvipsnames}{xcolor}
\usepackage{graphicx}
\usepackage{color}
\usepackage{amsmath}
\usepackage{amssymb}
\usepackage[utf8]{inputenc}
\usepackage{comment}
\usepackage{dcolumn}
\usepackage{bm}
\usepackage{hyperref}
\usepackage{soul}
\usepackage{footmisc}
\usepackage{xfrac}

\usepackage{ulem}
\usepackage{amsfonts}
\usepackage{verbatim}
\usepackage{stackrel}
\usepackage[capitalize]{cleveref}
\usepackage{tikz}
\usetikzlibrary{shapes.geometric, arrows}
\usetikzlibrary{decorations.markings}
\usetikzlibrary{decorations.pathmorphing}
\usetikzlibrary{decorations.pathreplacing}

\graphicspath{{/Images}}
\usepackage{float}
\newcolumntype{P}[1]{>{\centering\arraybackslash}p{#1}}
\definecolor{tdgreen}{rgb}{0,0.4,0}
\definecolor{tlgreen}{rgb}{0,0.7,0}
\definecolor{tpurple}{RGB}{103, 0, 31}
\definecolor{tbrown}{RGB}{127, 39, 4}
\definecolor{tviolet}{RGB}{73, 0, 106}
\definecolor{tgreen}{RGB}{1, 70, 54}
\definecolor{clblue}{rgb}{0.050, 0.35, 0.90}

\newcommand{\Rmnum}[1]{\expandafter\@slowromancap\romannumeral  #1@}

\usepackage{bbold}
\usepackage{color}

\makeatletter
\allowdisplaybreaks
\makeatother

\begin{document}

\title{Quantum Magnetic Skyrmions on Kondo-type Lattices}

\author{Zhen Zhao}
\email{Zhen.Zhao@teorfys.lu.se}
 \affiliation{Division of Mathematical Physics and ETSF, Lund University, PO Box 118, 221 00 Lund, Sweden}
 \author{Emil \"Ostberg}
 \email{Emil.Ostberg@teorfys.lu.se}
 \affiliation{Division of Mathematical Physics and ETSF, Lund University, PO Box 118, 221 00 Lund, Sweden}
 \author{Ferdi Aryasetiawan}
 \email{Ferdi.Aryasetiawan@teorfys.lu.se}
 \affiliation{Division of Mathematical Physics and ETSF, Lund University, PO Box 118, 221 00 Lund, Sweden}
 \affiliation{LINXS Institute of advanced Neutron and X-ray Science (LINXS), Mesongatan 4, 224 84 Lund, Sweden}
 \author{Claudio Verdozzi}
 \email{Claudio.Verdozzi@teorfys.lu.se}
 \affiliation{Division of Mathematical Physics and ETSF, Lund University, PO Box 118, 221 00 Lund, Sweden}
 \affiliation{LINXS Institute of advanced Neutron and X-ray Science (LINXS), Mesongatan 4, 224 84 Lund, Sweden}

\date{\today}

\begin{abstract}
A quantum description is given of nanoskyrmions in two-dimensional textures with localized spins and itinerant electrons, isolated or coupled to leads, in or out of  equilibrium. The spin-electron exchange is treated at the mean-field level, while tensor networks and exact diagonalization or nonequilibrium Green’s functions are used for localized spins and itinerant electrons. We motivate our scheme via exact and mean-field benchmarks, then show by several examples that itinerant electrons distinctly affect the properties of quantum nanoskyrmions. Finally, we mention lines of future work and improvement of the approach.
\end{abstract}

\maketitle
\textit{Introduction.} Magnetic skyrmions are topologically robust, long-lived spin textures  which can be manipulated by ultralow currents \citep{Bogdanov1994,Robler2006,
muhlbauer2009skyrmion,yu2010real,jonietz2010,
schulz2012,Huang2012,nagaosa2013,fert2013skyrmions,fert2017,Spaldin2022}, and of great potential for applications in spintronics and quantum computing \citep{zhang2020skyrmion,Picozzi2020,Yang2021,psaroudaki2021,Psaroudaki2023}. 
Skyrmions originate from the interplay of different types of magnetic interactions in solids \citep{szilva2023}, such as Heisenberg exchange \citep{heisenberg}, Dzyaloshinskii-Moriya interaction (DMI) \citep{dzyaloshinsky1958thermodynamic, moriya1960anisotropic}, and anisotropy \citep{Vaz_2008,Dieny2017}, to mention a few.

Theoretical descriptions of skyrmions are often based on classical spin-only models
\citep{Heinze2011,Makhfudz2012,Schutte2014,Psaroudaki2017}, treated via 
the Landau-Lifshitz-Gilbert \citep{Landau1935,Gilbert2004} and Thiele equations \citep{Thiele1973}.
These formulations, where itinerant electrons ({\it i} electrons) enter implicitly or as a source of renormalization of the classical spin dynamics, are in many cases adequate; however, describing explicitly the interaction (via Kondo-like exchange) between {\it i} electrons and localized spins  ({\it l} spins) can also be important \citep{bostrom2018,Nikolic1,Nikolic2,Bluegel2022,Bluegel2023,bostrom2022,Sahu2022,wang2023,ostberg2023,DMFT2023}, to, e.g., address interface phenomena like current driving and optical generation of nanoskyrmions \citep{Hellman2017}. 

In most of currently available {\it i} electrons+{\it l} spins schemes, 
spins are treated classically whereas electrons quantum mechanically
\citep{bostrom2018,Nikolic1,Nikolic2,Bluegel2022,Bluegel2023,bostrom2022,Sahu2022,wang2023,ostberg2023}, 
which makes computations viable for fairly large samples.
Such mixed quantum-classical strategy is reliable for skyrmion sizes of hundreds of lattice constants and/or large spin ($S \gtrsim 2$), where the role of quantum fluctuations is expected to be small.
However, in small systems, theoretical treatments based on classical and quantum spins can give different results (see e.g. Refs. \cite{mondal2021,Mazurenko2023}), suggesting that, to characterize experiments on nanometer-scale \cite{Heinze2011} skyrmions, a quantum description of spins \cite{ozawa2017,Sotnikov2021,Mazurenko2023,DMFT2023,haller2024}  is better suited.
Moreover, nanometer-scale skyrmions, due to their reduced spatial extent, exhibit pronounced susceptibility to boundary conditions, finite-size constraints, external field perturbations, and spin-transfer-induced phenomena such as spin slippage, all of which can critically influence the robustness of their topological protection.

For spin-only quantum descriptions, calculations are usually via exact diagonalization (ED) and tensor networks  (TNs) \citep{Cirac2021,lohani2019,haller2022,joshi2024quantum} (for a recent  analytical study, see  \citep{haller2024}). 
Another way to proceed is to treat the quantum skyrmion
Hamiltonian at the mean-field level \cite{Wieser_2017}. With this simplified yet computationally efficient approach, both equilibrium and real-time dynamics become tractable even for very large systems. The advantage over a fully classical approach lies in the inclusion of local spin quantum fluctuations, though the accuracy of the mean-field description compared to fully quantum skyrmion solutions remains essentially unexamined in general. 

When both localized spins and itinerant electrons are explicitly considered for skyrmion-type Hamiltonians
within a full quantum description, exact results are so far limited to very small systems, and primarily for benchmark uses; furthermore, treatments where the $l$ spin subsystem is treated at the mean-field level are also lacking.
It is thus highly desirable (in fact, necessary in several cases) to have a framework  including {\it i} electrons and 
{\it l} spin textures, and in the presence of reservoirs or leads, that at the same time does not neglect the quantum effects on nanoskyrmions. 

In a lattice-site representation, a common way to consider such interaction is to have at
each lattice site $R$ a local
Kondo-like exchange coupling between {\it i} electrons and {\it l} spins , i.e. $g \sum_R \hat{\bf {s}}_R^i \cdot \hat{\bf {S}}_R^l $. However, to proceed, approximations are needed, since in essence one is then dealing with a Kondo-lattice-like type of problem, further complicated by Heisenberg and DMI interactions for the {\it l} spins 
\footnote{In \citep{DMFT2023}, a two-band, real space DMFT approach to skyrmions was considered, with the dynamics treated in the linear response regime, and in the absence of electronic reservoirs.}. 

In this work, we propose a practically viable quantum treatment for both skyrmions and electrons, in and out of equilibrium. 
The approach uses an important level of approximation, namely a mean-field treatment of the Kondo-like exchange.
Nevertheless, it permits one to characterize intrinsically quantum physical quantities and gives access to
the full nonequilibrium dynamics (i.e., well beyond the linear-response regime), with electronic reservoirs included in the formulation. From now on, we will refer to this scheme
as QQmf, meaning that the we describe at the mean-field level the interaction between the two subsystems ({\it i} electrons and {\it l} spins) which are treated individually at the quantum level. Furthermore, we denote by QC the scheme with quantum {\it i} electrons and classical {\it l} spins, and by Wmf the scheme with quantum {\it i} electrons and quantum mean-field {\it l} spins.

In the rest of the Letter we exclusively consider the case of N\'{e}el-type skyrmions,
in situations where the Kondo-like coupling is ferromagnetic in character; 
the antiferromagnetic case is only briefly addressed in the Supplemental Material (SM) \footnote{See Supplemental Material
at [URL to be inserted] for a brief discussion of antiferromagnetic Kondo coupling. The Supplemental Material also includes computational details and additional benchmark comparisons.}. 

The main outcomes of our study are as follows:
(i) Exact numerical benchmarks show that our method has higher accuracy
over the QC  approach; 
this statement is supported by results  for both very small and larger systems, 
reported in the Letter and in the SM.
(ii) In many of the benchmark situations examined, the Wmf description improves over the QC one,
and is very close to the QQmf result. Still, for some physical quantities, a noticeable 
discrepancy remains between the Wmf and the QQmf, in both static and nonequilibrium
regimes.
(iii) The itinerant electrons
significantly influence a quantum skyrmion texture and its 
entanglement, quantum chirality, and quantum susceptibility properties.
(iv) In the dynamical regime following the application of a bias, 
quantum ``nanomerons" and bond currents dynamically affect each other in 
a highly nontrivial way at the atomistic lattice-site level.
These features are not accessible to spin-only descriptions and/or are described 
differently within QC and Wmf schemes.
%

\begin{figure}[t]
\centering
\includegraphics[scale=0.8] {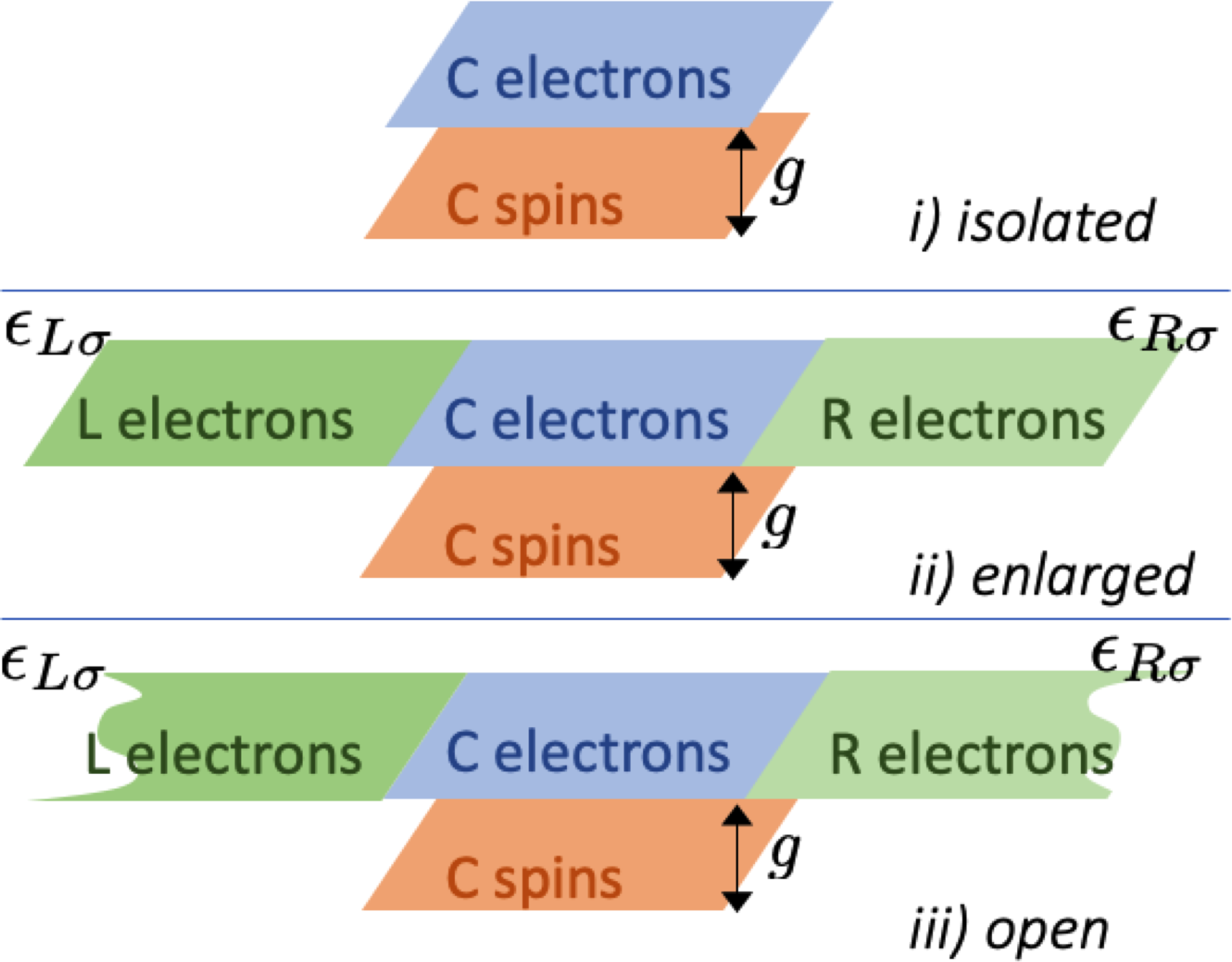}
\caption{Sketch of the setups considered, where  {\it i} electrons and {\it l} spins interact via local Kondo-like exchange of strength $g$.}
\label{fig:Comparison}
\end{figure}

\textit{System and Hamiltonian.}
As ``proof of concept'' of the method, we consider three setups of increasing complexity, shown in Fig.~\ref{fig:Comparison}. By denoting as $C$ ($L_E$) the regions where the {\it l} spins ({\it i} electrons) reside,  we have (i) both {\it l} spins and {\it i} electrons in the same finite lattice $C\equiv L_E$; (ii) {\it l} spins in $C$, and {\it i} electrons in an enlarged finite lattice $L_E=L+C+R$, where region $L$ ($R$) is coupled to $C$ from the left (right); (iii) same as (ii), but with semi-infinite $L$ and $R$ regions. We describe (i--iii) via a Kondo lattice Hamiltonian
plus Heisenberg exchange and DMI, i.e.,
$\hat{H}=\hat{H}^s+\hat{H}^d+\hat{H}^{sd}$ 
\footnote{
This Hamiltonian can be derived  from a two-band electron-only model, with electrons 
highly localized and correlated in one band (e.g., of the $d$ or $f$ type), and itinerant 
in the other band (e.g., of $s$ or $p$ type)  \citep{bostrom2022}}, with
\begin{eqnarray}
\!\!\!&&\hat{H}^s=-t_h\!\!\!\!\!\!\sum_{\langle ij \rangle\in L_E;\sigma}\!\!\!\big(\hat{c}_{i\sigma}^\dagger \hat{c}_{j\sigma} + \text{H.c.}\big), \quad
\hat{H}^{sd}=-g\sum_{i\in C} \hat{\mathbf{s}}_i\cdot \hat{\mathbf{S}}_i\nonumber\\ 
\!\!\!&&\hat{H}^d=\!\!\!\!\!\!\sum_{\langle ij \rangle\in C}\Big[\mathbf{D}_{ij}\cdot\hat{\mathbf{S}}_i\times\hat{\mathbf{S}}_j-J_{ij}\hat{\mathbf{S}}_i\cdot\hat{\mathbf{S}}_j\Big]-\sum_{i\in C}\mathbf{B}_i\cdot\hat{\mathbf{S}}_i\label{eq:Hamiltonian}.
\end{eqnarray}
Here, $\hat{H}^{s}$ is the {\it i} electrons part, with hopping parameter $t_h$ between nearest-neighbor (n.n.) sites, and
$\hat{c}_{i\sigma}^\dagger$ creating an electron with spin projection $\sigma$ at site $i$. The {\it l} spins texture
is described by $\hat{H}^{d}$, where $\mathbf{\hat{S}}_i$ is the spin operator at site $i$, and  
$J$, $\mathbf{B}$, and ${\bf D}$, respectively, denote the Heisenberg exchange, the 
external magnetic field, and the N\'{e}el-type DMI, namely, ${\bf D}_{ij}=D\frac{{\bf r}_{i}-{\bf r}_{j}} {| {\bf r}_{i}-{\bf r}_{j}|}\wedge \hat{\bf z}$.
Finally, $\hat{H}^{sd}$ is the  Kondo-like exchange term , with $\mathbf{\hat{s}}_i=(1/2)\sum_{\tau \tau'} \hat{c}^\dagger_{i\tau} \boldsymbol{\sigma}_{\tau\tau'} \hat{c}_{i\tau'}$ and 
$\boldsymbol{\sigma}\equiv (\sigma^{x},\sigma^{y},\sigma^{z})$ the vector of Pauli matrices. 
In $\hat{H}^s$, there is no spin-orbit interaction (SOI). Usually small, SOI can sometimes affect skyrmion stabilization \citep{Manchon2015,Hellman2017,Rowland2016,Sahu2022}), but it is not considered here to limit the number of model parameters.
%
\begin{figure}[t]
\centering
\includegraphics[scale=0.55] {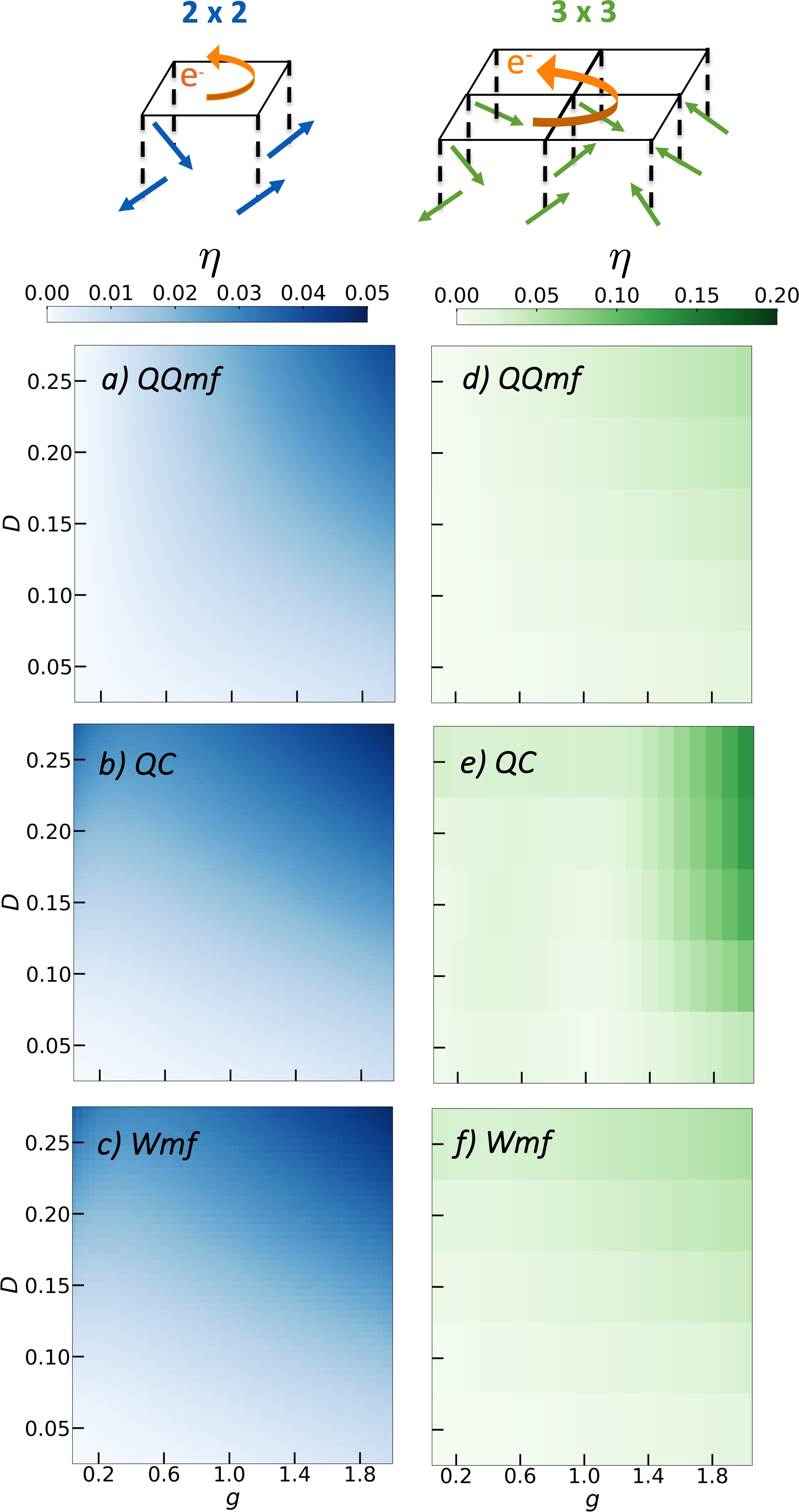}
\caption{Relative average error $\eta$ [Eq.~\ref{Eq4eta}] between the proposed QQmf scheme with respect to
the full TN solution for a 2$\times$2 (left) and a 3$\times$3 (right) plaquette. Results for
$\eta$ are also shown for quantum electron+classical spin (QC) solution and for the 
case where the localized spins are treated at the quantum mean-field level (Wmf).
The parameters used are 
$J\!=\!0.1,B^z\!=\!0.1,t_h\!=\!1$. For the case $J=1$, see the main text and the SM.}
\label{fig:NewComparison}
\end{figure}

\textit{Approach and method of solution.} 
We will treat $\hat{H}^{sd}$ at the mean-field level.
Hence, by defining $\hat{V}^s=\sum_{i\in C} \ex{\hat{\mathbf{S}}_i}\cdot \hat{\mathbf{s}}_i$ and  $ \hat{V}^d=\sum_{i\in C} \ex{\hat{\mathbf{s}}_i}\cdot \hat{\mathbf{S}}_i$, 
we have $\hat{H}\rightarrow  (\hat{H}^s_\text{MF}, \hat{H}^d_\text{MF})$, with
\begin{eqnarray}
\label{Hamiltonian_mean}
\hat{H}^{s/d}_\text{MF}=\hat{H}^{s/d}-g\hat{V}^{s/d}.
\end{eqnarray}
The results in the paper are obtained using Eq.~(\ref{Hamiltonian_mean}).\\
{\it Isolated or enlarged system.} Starting with $\ex{\mathbf{s}_i}=0$, we use a
matrix product state (MPS) algorithm \citep{Fannes1992,ostlund1995,rommer1997} 
from the ITensor library \citep{itensor} to find the ground state of $\hat{H}^d_\text{MF}$ and update the averages $\ex{\mathbf{S}_i}$. These then enter as parameters when solving via ED for $\hat{H}^{s}_\text{MF}$, while reproducing $\ex{\mathbf{s}_i}$. As the iterations converge, the mean-field ground state is reached (see the Supplemental Material (SM) for details). For the enlarged case, the procedure is carried out with $\epsilon_{i\sigma}=0$ at $t=0$. The time evolution for the enlarged system is discussed later in the Letter. \\
{\it Open system.} Starting with the ground state of the isolated region $C$, 
the tunneling matrix elements between $C$ and the leads $R$, $L$ are
switched on in time adiabatically up to a final value, so that $L+C+R$ reaches a steady state (chosen as the initial or ground state of the open system). In the process, the quantum {\it l} spins evolve in time via time-evolution block-decimation (TEBD) \citep{PAECKEL2019}, and the {\it l}-electrons via the nonequilibrium Green's function (NEGF) approach \cite{KadanoffBaym,keldysh1965diagram,bonitz2012,stefanucci2013} in the equal-time formulation. In this scheme, the key quantity is the one-particle density matrix $\rho$ for region C, related to the full one-particle NEGF $G(t,t')$ via 
$\rho(t)=-iG(t,t^+)$. With leads, $\rho$ obeys 
\begin{eqnarray}
\label{eq:rho_eom}
i\frac{d}{dt}\rho(t)=[h(t)\rho(t)-iI_\text{col.}(t)]-\text{H.c.}
\end{eqnarray}
Here, $h(t)$ is the (possibly time-dependent) single-particle term and 
$I_\text{col.}$ is the collision integral \citep{Latini2014},
that in general describes interaction-induced correlation effects and lead-induced embedding effects. The {\it i} electrons do not interact in our model; i.e., $I_\text{col.}$ has only the embedding part. We also (i) use the so-called wide-band limit (WBL), where the hopping term $t_l$ in $L,R$ tends to infinity, while $\gamma= t_l^2/t_h$ remains fixed ($\gamma$ thus sets the magnitude of the $L$-$C$ and $R$-$C$ couplings), and
(ii) further approximate the WBL by simplifying the calculation of $I_\text{col.}$. Steps (i,ii) are a good trade-off between reduced computation time and accuracy \citep{ostberg2023} (see the SM).\\
{\it Mean-field and classical schemes for the localized spins.} As terms of comparison to the 
QQmf method, we considered (i) quantum mean-field and (ii) classical descriptions of the spin texture, while keeping the electrons and the Kondo-like
coupling treated as in the QQmf scheme. Description (i) is  
based on the Weiss-like mean-field (Wmf) approach taken from 
\cite{Wieser_2017} and briefly summarized in the SM; details
of (ii) can be found in, e.g., \cite{bostrom2018,ostberg2023}.\\
%
\begin{figure*}[t]
\centering
\includegraphics[width=\linewidth]{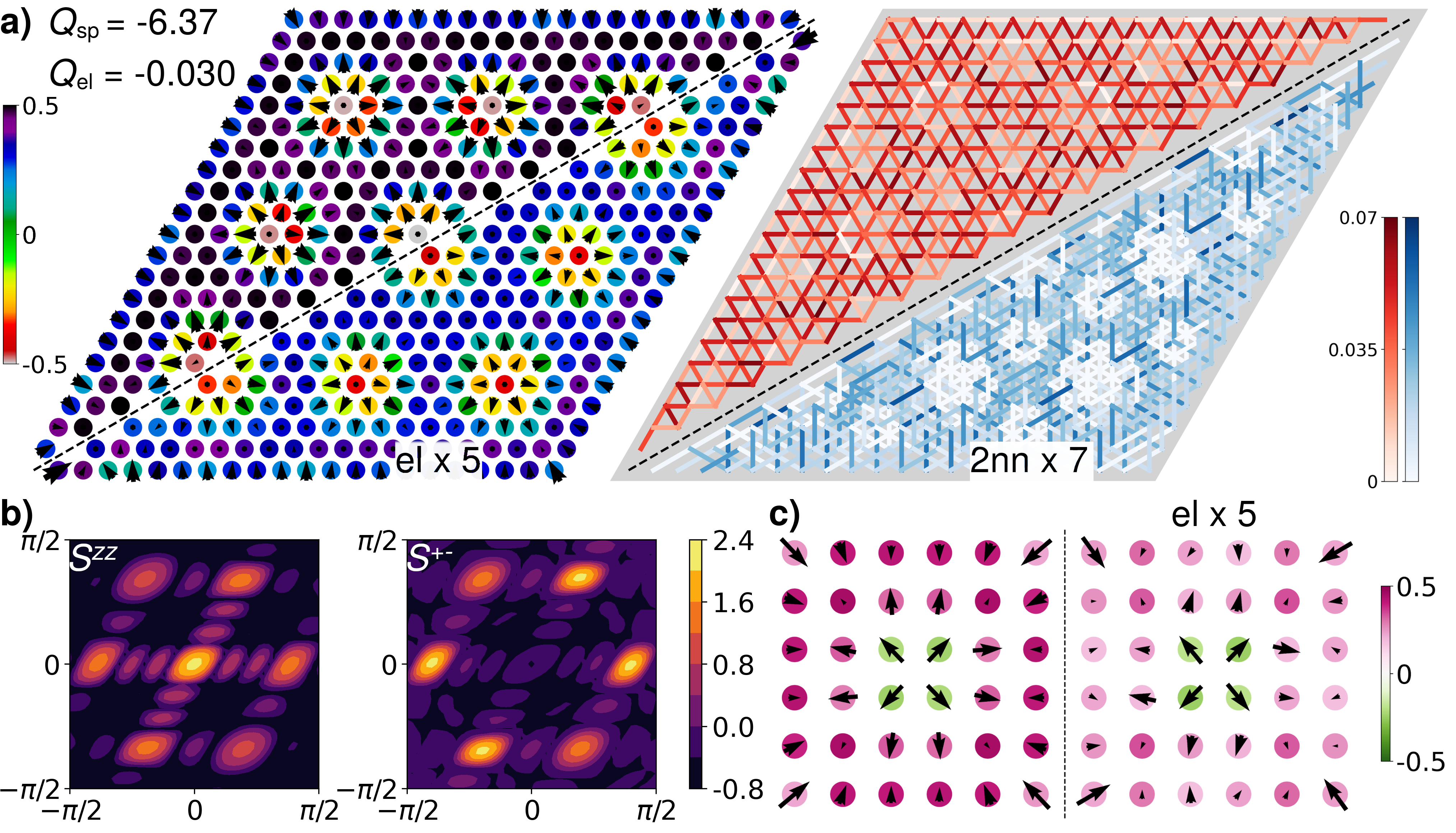}
\caption{Ground state results for (a), (b) an isolated system with $D=0.2,J=0.1,B^z=0.1,t_h=1,g=2$, and (c): an open system with $D=1,J=0.2,B^z=0.5,t_h=1,g=1$.
(a) Spin expectation value (left) and
concurrences  $\mathcal{C}$ (right) heat maps in a $21\times21$ isolated rhombus cluster, with {\it l} spin ({\it i} electron) chiralities $Q_\text{sp}=-6.37$ ($Q_\text{el}=-0.03$). 
In the spin map, the top (bottom) half shows the expectation value $\ex{S^{x,y,z}}$ ($\ex{s^{x,y,z}}$) of the {\it l} spins (the {\it i}electrons). The values of the $z$- and $xy$ spin-components are indicated by colours and arrows,
respectively. In the concurrence map, the value of $\mathcal{C}$ for n.n. (upper half) and next-nearest-neighbour (2n.n., lower half) sites is represented by the bond color. The missing half of each map is recovered using rhombus symmetry. 
(b) Logarithm of the static structure factors $\ln S^{zz}$ (left)  and $\ln S^{+-}$ (right) of the {\it l} spins.
(c) $\ex{S^{x,y,z}}$ (left half) and $\ex{s^{x,y,z}}$ for a $6\times6$ square cluster contacted to leads, with parameters $D=1,J=0.2,B^z=0.5,t_h=1,g=1$.}
\label{fig:21}
\end{figure*}
%
%
\begin{figure}[t]
\centering
\includegraphics[width=\linewidth]{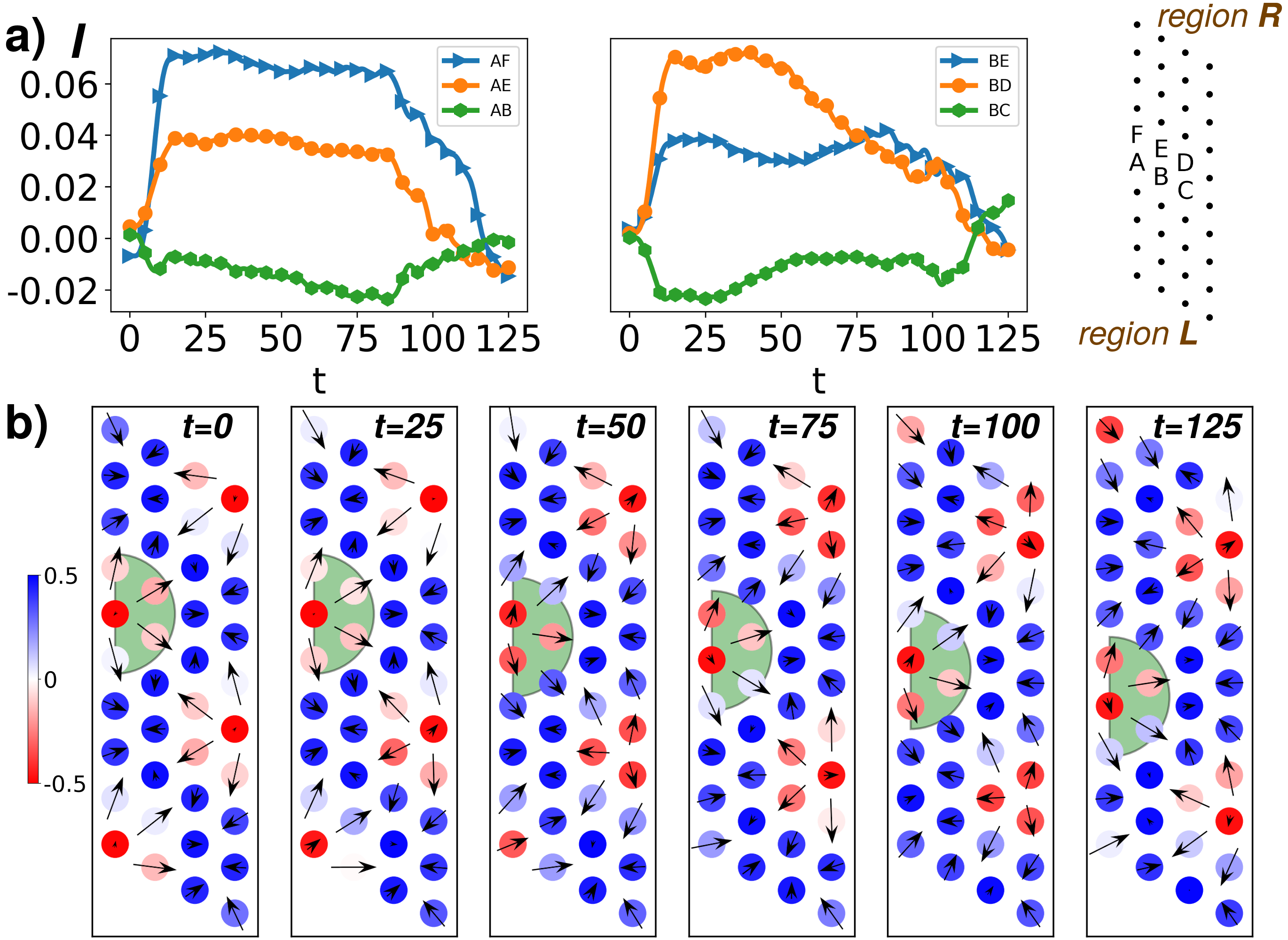}
\caption{Nonequilibrium dynamics of a rhombus cluster $C$ of $4\times10$ sites with an enlarged electron region $L+C+R$ of $4\times(100+10+100)$ sites, for parameters $D\!=\!0.2, J\!=\!0.05, B^z\!=\!0.06, t_h\!=\!1, g\!=\!2$. The dynamics is induced by a spin-up polarized bias $\epsilon_{L}=-\epsilon_{R}$ in the enlarged region, switched on at $t\!=\!0$ and ramped to maximum strength 0.5 at $\!t=\!10$. (a) Spin-up current  $I_{\ex{ij}}$ along the bonds $\langle ij\rangle$ labeled by A--F (sketch on the right). (b) Time snapshots of the spin expectation values $\ex{S^{x,y,z}(t)}$ in $C$, $z$- and $xy$-spin components indicated by colors and arrows, respectively.
The current-driven motion of one meron is highlighted by the green semicircle. }
\label{fig:cur}
\end{figure}
\textit{Exact versus mean-field Kondo-like exchange, and quantum versus classical spins.}
To see if a quantum treatment of {\it l} spins+{\it i} electrons (albeit within a mean-field account of the Kondo-like exchange) is beneficial, we consider 2$\times$2  and 3$\times$3  square 
plaquettes (respectively, with 4 spins + 4 electrons and  9 spins + 9 electrons) described by
Eq.~\eqref{eq:Hamiltonian}, and with open boundary conditions (OBC). Via the TN method, we determine the
full quantum ground state of these systems, which is then used as a reference to assess the scope of the 
QQmf and QC descriptions. We do this via the quantity (A = QQmf or QC)
\begin{equation}
 \eta_\text{A}=\frac{1}{N}\sum_{i} \frac{ |\langle {\hat{\bm S}_i} \rangle_\text{A} - \langle {\hat{\bm S}_i}\rangle_\text{TN} |}{ |\langle {\hat{\bm S}_i} \rangle_\text{TN}|},
 \label{Eq4eta}
\end{equation} 
The indication from panels (a,b) of Fig.~\ref{fig:NewComparison} is that QQmf and QC give similar and
overall satisfactory results for the 2$\times$2 system, although the QC description is slightly inferior in a subregion of the heat-map. On the other hand, for the 3$\times$3 plaquette [panels (c,d)], the QQMF solution is markedly better, especially in the region corresponding to large $g$ and $D$ (for reference, the 
QQmf treatment is exact for $g=0$, as reflected in the heat maps). While providing limited evidence, the results of Fig.~\ref{fig:NewComparison} could also hint that the difference in quality of the QQmf and QC treatments may increase for larger system size, thus lending support to the choice of parameter values considered in the rest of the Letter.

Besides the results of Fig.~\ref{fig:NewComparison}, in the SM we report
several other comparisons in terms of the 2$\times$2  and 3$\times$3 systems, and for 
both ferromagnetic (FM) ($g>0$) and antiferromagnetic (AF) ($g<0$) couplings between {\it i} electrons and {\it l} spins.
The overall picture emerging from Fig.~\ref{fig:NewComparison} and
the additional testing is that for the FM case there is
an appreciable region of the $g, D$ values where the QQmf approach performs better than the QC treatment,
while at small $g, D$ values QQmf and QC retain a similar (good) performance.
This behavior occurs for either small or large values of the FM coupling among $\it l$ spins. On the
other hand, in the AF case, both QQmf and QC do not perform well already for rather moderate 
$g, D$ values, due to the inadequacy of the mean-field treatment in the Kondo regime.
 
Next, to discuss the role of the {\it i} electrons, we move to a spin-electron (sp-el) 
and a spin-only (sp) system in a rhombus-shaped
$11\times11$ finite triangular lattice, also treated with OBC. We again compare their respective ground states
using $\eta'=N^{-1}\sum_{i\in C} \big |\ex{\hat{\mathbf{S}}_i^\text{sp-el}}-\ex{\hat{\mathbf{S}}_i^\text{sp}}\big |\big/\big |\ex{\hat{\mathbf{S}}_i^\text{sp}}\big|$. We choose $D\!\!=\!\!0.2, J\!\!=\!\!0.15, B^z\!\!=\!\!0.06, t_h\!\!=\!\!1, g\!\!=\!\!2$, for which
both systems exhibit a single-skyrmion texture. This is confirmed by the value of 
the quantum scalar chirality \citep{Sotnikov2021} 
$Q=\pi^{-1}\sum_{<ijk>\in C}\ex{\hat{\mathbf{S}}_i\cdot[\hat{\mathbf{S}}_j\times \hat{\mathbf{S}}_k]}$, where the sum runs over all nonoverlapping triangles formed by neighboring sites $i$, $j$, $k$. For $S=1/2$ {\it l} spins, $|Q|\sim 1$ signals the presence of a skyrmion. For the case at hand
we respectively find $Q_\text{sp}^\text{sp-el}=-0.618$, and $Q^\text{sp}=-0.624$. Even in this strong Kondo regime ($g/t_h=2$),  the mean-field electron-spin expectation values $\ex{\hat{\mathbf{s}}_i}$ are only a few percent of the corresponding $\ex{\hat{\mathbf{S}}_i^\text{sp-el}}$. Yet, since $\eta'=0.14$, the sp-el and sp ground states noticeably differ. 
Altogether, these comparisons confirm that {\it i} electrons markedly affect the quantum nanoskyrmion and that it is beneficial, even with a mean-field treatment of Kondo-like exchange, to have quantum rather than classical {\it l} spins. 


\textit{Skyrmion ground states.} To simulate nanosized skyrmions in isolated or open systems we take $D/J\sim 2-5$. These are
large values, but consistent with the giant DMI ones from some monolayer materials \citep{Zhang2023}. The other parameters used, $J/t_h\sim0.1-0.5$ and $g/t_h\sim1-2$, are also congruous with typical values in the literature \citep{mondal2021}.\\
\textit{Isolated system.} We consider a $21\times21$ rhombus cluster (which permits to simulate a skyrmion lattice) with OBC and parameters $D=0.2,J=0.1,B^z=0.1,t_h=1,g=2$. The ground state profile of such spin+electron system is shown in Fig.~\ref{fig:21}(a), where the skyrmions exhibit an
approximate periodic alignment, while the sites 
with  $S^z \simeq \frac{1}{2}$ state (black circles) can be thought of as domain walls. 
Since the {\it l} spins are described quantum mechanically, we can look  
at entanglement between spin pairs at sites $i\neq j$ \citep{Wootters1998,horodecki2009,haller2022}, quantified by the concurrence $\mathcal{C}_{ij}=\mathcal{C}[\hat{\rho}_{ij}]$, and where
 $\hat{\rho}_{ij}$ is the reduced density matrix. Similarly to what is
found in \citep{haller2022}, $\mathcal{C}$  shows a lack of long-range entanglement for the skyrmion (crystal) texture. However, since the distance $\eta'$ between sp-el and sp systems is $10\%$, 
the concurrences with and without {\it i} electrons discernibly differ. 
In turn, the {\it i} electrons are affected by the skyrmion texture (for example,
the local density of states for spin-up and spin-down {\it i} electrons
show large (small) imbalance at (far away from) the skyrmions-core sites, 
see the SM).

Further insight comes from the spin structure factor, defined as $S^{\alpha\beta}(\mathbf{k})=N^{-1}\sum_{ij\in C}\ex{\hat{S}_i^\alpha\hat{S}_j^\beta}e^{i\mathbf{k}\cdot(\mathbf{r}_i-\mathbf{r}_j)}$, and with $\alpha\beta$ denoting either the $zz$ or the $+-$ components. In Fig.~\ref{fig:21}(b) (for clarity, we show $\ln S^{\alpha\beta}$ rather than $S^{\alpha\beta}$), both $S^{zz}({\bf k})$ and $S^{+-}({\bf k})$ exhibit
sixfold intensity patterns, consistent with neutron scattering results (from e.g. the B20 compound MnSi \citep{muhlbauer2009skyrmion}). Overall, the signal is stronger for $S^{zz}$ than for $S^{+-}$; in particular, $S^{zz}$ has a bright spot at ${\bf k}=0$, due to the
strong degree of spin orientation along $z$. Additionally, the logarithmic display
reveals additional but much less intense, lower symmetry features, that we ascribe to size 
and the open-boundary-conditions effects in the rhombus cluster.

\noindent \textit{Ground state for an open system.}
Using NEGF for the {\it i} electrons and tensor networks for the {\it l} spins, 
we have also considered the case of a $6\times6$ sp-el square cluster 
attached to leads. The isolated central region ground state at $t=0$ (determined in the mean-field Kondo-like exchange) is connected to the leads adiabatically for $0<t<50$. The evolution of the sp-el system can then be continued to $t_0=125$, which establishes steady spin-polarized currents through the central region in equilibrium. These currents are weak and essentially present only at the boundary sites of $C$ (see the SM).  
The ground state average spin distributions for this setup, with $D=1,J=0.2,B^z=0.5,t_h=1,g=1$, are shown in Fig.~\ref{fig:21}(c). They result in a value
$\eta=0.06$, i.e., a nonstrong average influence of the electrons on the spin texture.  
Still, an appreciable and interesting feature is that the spins of the {\it i}-electrons are tilted compared to the isolated case (not shown). Since for the open case there is a spin-polarized electric current flowing (and a corresponding induced magnetic field), we attribute the spin tilt to an induced effective spin-orbit interaction.

\noindent \textit{Dynamics.}
For out-of-equilibrium situations, the natural option would be to consider the open system of 
Fig.~\ref{fig:Comparison}iii where, after reaching equilibrium at $t=t_0$ (as discussed above,
via a slow-ramping dynamics to connect the leads and dissipate fluctuations), 
time-dependent currents are injected in the electron subsystem via a bias. For example,
for spin-up polarization and with reference to Eq.~(\ref{Hamiltonian_mean}), 
$\hat{H}^{s}\rightarrow \hat{H}^{s}(t)\equiv \hat{H}^{s}+\hat{w}_L(t)+\hat{w}_R(t)$, where
$\hat{w}_{L}(t)=\sum_{i\in L} \epsilon_{L}(t)\hat{c}^\dagger_{i\uparrow}\hat{c}_{i\uparrow}$ and  similar for 
$\hat{w}_{R}(t)$.
The time evolution would then be performed via NEGF for the electrons and via, e.g., TEBD for the spins.

We defer the study of the NEGF+TEBD dynamics 
of a {\it l} spins+{\it i} electrons system connected to semi-infinite leads to a forthcoming paper. Here, as proof of concept of our approach, we use the enlarged isolated setup of Fig.~\ref{fig:NewComparison}(b), with a region $C$ consisting of a $4\times10$ rhombus cluster, and an enlarged electron region of $4\times(100+10+100)$ sites, to delay the reflection of the currents by the boundaries, and to have steady, stable currents established within the time of interest. The system is considered within OBC.
In our simulations, we set $t_0\!=\!0, D\!=\!0.2,J\!=\!0.05, B^z\!=\!0.06, t_h\!=\!1, g\!=\!2$.
In this case, the QQmf ground state $|\Psi^\text{el}(0)\rangle$ [obtained without
time propagation,  
and shown at $t\!=\!0$
in Fig.~\ref{fig:cur}(b)] corresponds to a spin texture with multiple meron-like (i.e. half-skyrmion-like) structures. These
(henceforth called merons for simplicity) are centered at sites $i$ of the cluster boundaries where $\ex{\hat{S}_i^z}<0$, with
$\ex{\hat{S}_{i'}^z}>0$ for their second 
and third neighbors' sites $i'$. 
To support this interpretation and address  the individual behavior of merons, we use a modified, ``local'' scalar chirality $Q_i$ calculated over all the triangles
including sites $i$ and $i'$ (for example, at $t\!=\!0$, $Q_i \sim -0.16$ for the green shaded meron in Fig.~\ref{fig:cur}(b);
see the SM for details).

Then, starting from $|\Psi^\text{el}(0)\rangle$, we introduce for $t>0$ a bias  
$\epsilon_{L}(t)=0.5[1-\theta(10-t)\cos^2(\pi t/20)]$
with $\epsilon_{R}(t)=-\epsilon_{L}(t)$.  At each time step of the time evolution, $\hat{H}^d_\text{MF}$ is constructed with input from the time-evolved spin expectation $\ex{\hat{\mathbf{s}}_i}_t$, and the {\it l} spins are evolved via TEBD. Subsequently, an updated $\ex{\hat{\mathbf{S}}_i}_t$ and the bias $\epsilon_{L/R}(t)$ drive the electrons via $\hat{H}^s_\text{MF}(t)$, and the full spin-electron system is thus evolved for one time step. 

\begin{figure}[t]
\centering
\includegraphics[width=\linewidth]{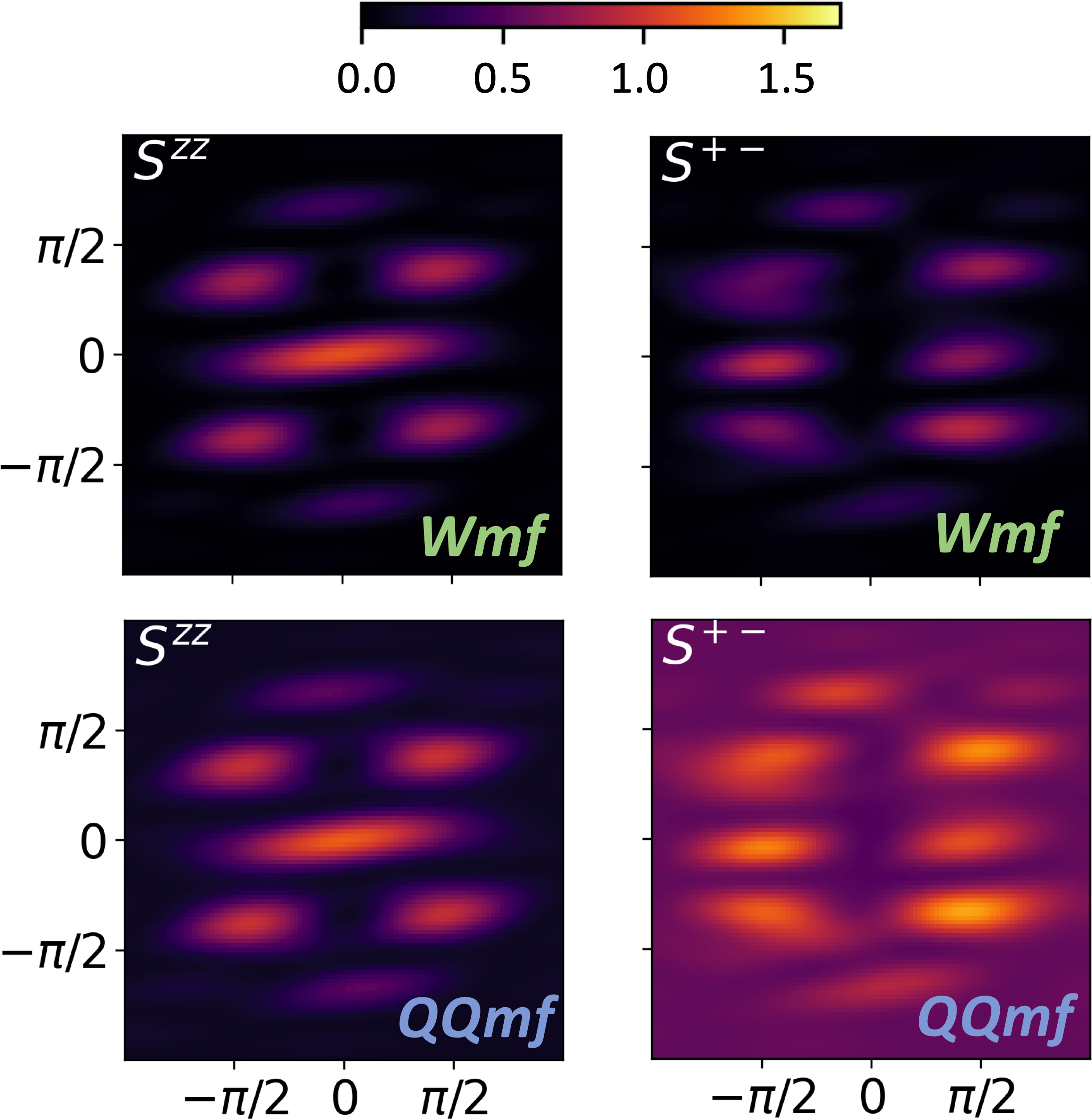}
\caption{Wmf (top) vs QQmf (bottom) structure factor $S^{\alpha\beta}(\mathbf{k})$ at $t=125$
for the  $4\times(100+10+100)$ $i$ electron+$l$ spin texture. Model parameters are $D=0.2,J=0.05,B^z=0.06,t_h=1,g=2$. Spin-up polarized bias $\epsilon_{L}=-\epsilon_{R}$ in the enlarged region switched on at $t\!=\!0$ and ramped to maximum strength 0.5 at $\!t=\!10$.}
\label{fig:additional}
\end{figure}

The time evolution of selected spin-up bond currents 
$I_{\ex{ij}}=-2t_h\text{Im}\ex{\hat{c}_{j\uparrow}^\dagger \hat{c _{i\uparrow}}}$
is shown in Fig.~\ref{fig:cur}(a). 
After the transient regime ($t \lesssim 10$), the currents at or near the edges ($I_{AF, AE, AB}$) stay rather stable until $t\approx 80$; then, they start to be reflected at the outer boundaries of the $L, R$ regions. By contrast, currents away from the edges ($I_{BE, BD, BG}$) undergo reflection sooner (i.e., $t \approx 50$). In Fig.~\ref{fig:cur}(a), the currents at different bonds differ for magnitude and/or sign (also for parallel adjacent bonds). This is because of the merons: these are driven in a direction opposite (see the shaded area in the time snapshots) to the average flow of the spin-up electrons
(also, small size and nonequilibrium quantum fluctuations make the merons slightly change their shape in time).
The {\it i} electrons, in turn, conform their flow to the presence of the moving merons, which results in
the observed behavior of the bond spin-up currents. Furthermore, in the ground state ($t=0$),
the {\it i} electron's spins texture mirrors the {\it l} spins one, but in a much weaker pattern.
The latter extends outside region $C$, while fading as the distance from $C$ increases. 
Yet, with currents (i.e. $t>0$), the outer parts of the pattern move towards $C$, 
affecting the {\it l} spins texture.
This is why, for example, $\ex{\hat{S}^z}$ at the top-left site in Fig.~\ref{fig:cur}(b) changes 
sign during the time evolution.
To further qualify the QQmf approach and the inherent results, it can be useful to employ the Wmf approach for the present system and situation. In doing so, we find that the local spin expectation values from the QQmf and Wmf methods are initially quite similar [$\eta(t=0)=3$], but their difference increases during time evolution [$\eta(t=125)=7$]. 
We also examined the spin-spin correlation functions $S^{\alpha\beta}(\mathbf{k})$ (within Wmf,
$\langle \hat{S}^\alpha_i \hat{S}^\beta_j\rangle\approx \langle \hat{S}^\alpha_i \rangle \langle \hat{S}^\beta_j\rangle$).
These look different in the two descriptions, as illustrated for $S^{\alpha\beta}[\mathbf{k}(t=125)]$ 
in Fig.~\ref{fig:additional} (and for $S^{\alpha\beta}[\mathbf{k}(t=0)]$ in the SM). 
These trends align with the typical behavior of mean-field treatments in condensed matter: 
while corrections to mean-field values may often be small for some observables, they can still produce noticeable 
differences in some physical quantities.

\textit{Conclusions.} We have proposed a computationally viable theoretical approach to describe 
quantum nanoskyrmions for in- and out-of-equilibrium itinerant-electron+localized-spin systems, with 
scope and results different from classical-spin+quantum-electron or spin-only treatments. 
It combines matrix product states methods for the localized-spins, with exact diagonalization or nonequilibrium 
Green's function methods for the itinerant electrons, while treating the spin-electron interaction at the mean-field level. 

Benchmark comparisons for both small and larger systems show that the newly introduced 
QQmf method surpasses the quantum-classical (QC) approach in terms of accuracy. At the same time, 
the quantum mean-field treatment of the localized spins (Wmf) not only improves upon the QC approach but also, 
in some cases, closely matches the results of the more sophisticated QQmf technique. However, noticeable 
differences can still emerge between the Wmf and QQmf descriptions, in both static and dynamic conditions.

Besides nanoskyrmions, the method permits one to address a vast range of physical phenomena in (topological or not, 
ordered or not) magnetic systems 
that are elusive to classical-spin or spin-only schemes, for example, magnetoresistance due to quantum spin textures, the photoexcitation of quantum skyrmions, or the photoelectron spectrum from the itinerant electrons in the presence of quantum spin textures. Furthermore, the scheme is amenable to several extensions, and thus likely expected
to foster new avenues of research. As next developments, we are considering ways to go beyond a mean-field treatment of the spin-electron interaction via nonperturbative local approximations, and the inclusion photon-skyrmion interactions via tensor networks \citep{Huggins_2019,Shinaoka2023} and nonequilibrium Green's function formulations \cite{Pavlyukh1,bostrom2022,Megha2023}.\\

\begin{acknowledgments}
Z.Z., E.Ö., and C.V. gratefully acknowledge financial support from the Swedish Research Council 
(Vetenskapsrådet, VR, Grant No. 2017-03945 and No. 2022-04486).
F.A. gratefully acknowledges
financial support from the Swedish Research Council 
(Vetenskapsrådet, VR, Grant No. 2021-04498\_3). 
\end{acknowledgments}

\bibliography{bib1}

\end{document}